\documentclass[submission,copyright,creativecommons]{eptcs}
\usepackage{xcolor}
\usepackage{hyperref}
\usepackage{listings}
\usepackage[dvips]{graphicx}
\usepackage{subcaption}
\usepackage{amsthm}
\usepackage{enumitem}
\usepackage{amsmath}
\usepackage{amsfonts}
\usepackage{cleveref}

\theoremstyle{definition}
\newtheorem{example}{Example}

\providecommand{\event}{FROM 2025 (Working Formal Methods Symposium)}

\definecolor{codegreen}{rgb}{0,0.6,0}
\definecolor{codegray}{rgb}{0.5,0.5,0.5}
\definecolor{codepurple}{rgb}{0.58,0,0.82}
\definecolor{backcolour}{rgb}{0.95,0.95,0.92}

\usepackage{iftex}

\graphicspath{{images/}}

\ifpdf
  \usepackage{underscore}         
  \usepackage[T1]{fontenc}        
\else
  \usepackage{breakurl}           
\fi

\title{Navigating the Python Type Jungle}

\author{
    Andrei Nacu
    \institute{Faculty of Computer Science\\
    Alexandru Ioan Cuza University\\
    Iasi, Romania}
    \email{andreinaku@gmail.com}
    \and
    Dorel Lucanu
    \institute{Faculty of Computer Science\\
    Alexandru Ioan Cuza University\\
    Iasi, Romania}
    \email{Dorel.Lucanu@gmail.com}
}

\def\titlerunning{Navigating the Python Type Jungle}
\def\authorrunning{A. Nacu, D. Lucanu}
\begin{document}
\maketitle

\begin{abstract}
Python's typing system has evolved pragmatically into a powerful but theoretically fragmented system, with scattered specifications. This paper proposes a formalization to address this fragmentation. The central contribution is a formal foundation that uses concepts from type theory to demonstrate that Python's type system can be elegantly described. This work aims to serve as a crucial first step toward the future development of type inference tools.
\end{abstract}

\section{Introduction}
The evolution of Python's typing system has been pragmatic, driven by the practical needs of programmers. This has resulted in a flexible and powerful system. However, its official specification is scattered across Python Enhancement Proposals (PEPs) and module documentation. As a consequence, a holistic theoretical understanding of the type system is challenging. This paper aims to bridge this gap by establishing a formal theoretical foundation for Python's type model.

The central contribution of this work is to provide a different lens with which to look at Python types. We believe that the Python typing system, originally designed by Guido van Rossum, is so powerful that it can be elegantly described using concepts such as abstract data types (ADTs) and existential types~\cite{mitchellexistential}. 
The following separately known facts embody the main idea:
\begin{quote}
    Every Python type is represented by a class.\\
    A class is an implementation of an abstract data type.\\
    An abstract data type has an existential type.
\end{quote}
The proposed formalisms are also intended to be used as a foundation for a type inference framework that aims to compute the possible types of classes and functions in isolation.
To our knowledge, this is the first place where the existential types are used as an unifying framework to formalize the Python type-trelated concepts.

The remainder of this paper is organized as follows: in the second section, we explain crucial concepts, such as runtime classes, type annotations, emphasizing abstract base classes and protocols. These are important because they are widely used in stub files that describe specifications for built-in and popular third party Python modules. In the third section, we lay the groundwork for a formalization of types based on ADTs and existential types. We start out by explaining the foundational concepts of our formalization and finish by applying these concepts to describe Python typing concepts. The fourth section contains related work that summarizes research papers, specifications and static type checker applications which will be of aid in our process. Finally, we conclude by summarizing our findings and outline future objectives that help us achieve a sound formalization of Python types.

\section{Type Related Concepts Used in Python}

\subsection{Brief History}
\label{subsec:history}
A cornerstone principle of Python is that \emph{everything is an object}. This includes classes as well. While this principle has been constant since Python's inception, its implementation has evolved. Python 2.2 introduced a significant architectural change. This version triggered the process of \emph{type/class unification}~\cite{pyunif}, in which distinctions between built-in types and user-defined classes were eliminated. This version introduced new-style classes, differentiating them from the legacy old-style ones. Each new-style class inherits from the same class, \texttt{object}. However, the problem that the method resolution order algorithm (MRO) had inconsistent behavior remained. Python 2.3 further refined this evolution by introducing C3~\cite{c3paper} as the MRO algorithm~\cite{python23c3}. This provided a deterministic, robust and predictable way to linearize complex inheritance hierarchies, making Python's object model more reliable. Another notable development in this process occured with the release of Python 3, which deprecated old-style classes.

It is fair to ask why this change was needed, or what benefits it brought. Before the introduction of new-style classes, Python had two distinct types of classes: built-in and user-defined. Built-in types were implemented in C and exposed to Python, while user-defined classes were implemented in Python itself. Built-in classes represented the core data types of Python, like integers, strings, lists, etc. A major problem with this model was that built-in classes were not flexible enough to be subclassed or extended by user-defined classes, thereby making the creation of custom data types difficult. The class unification process merged both types of classes into a single class hierarchy. Therefore, built-in classes and user-defined classes inherited from the \texttt{object} class and effectively became new-style classes.

To define a new-style class, the user would define a class that would inherit from \texttt{object} (or any other new-style class).
\begin{lstlisting}[language=Python, numbers=none, caption=Python 2.2 class definition]
class foo(object):  # new-style class
    pass

class bar(foo):  # also a new-style class
    pass

class baz:  # old-style class
    pass
\end{lstlisting}
An integral part of the unification process was to make built-in classes inherently new-style. Afterward, Python 3 eliminated old-style class support by making classes new-style by default.
\begin{lstlisting}[language=Python, numbers=none, caption=Python 3 class definition]
class foo:  # new-style class
    pass

class bar(foo):  # also a new-style class
    pass

class baz:  # new-style class
    pass
\end{lstlisting}

\subsection{Core Typing Concepts}

\subsubsection{Metaclasses}
\label{subsec:metaclasses}
Since everything is an object in Python, classes are no exception. A class that is responsible for defining and constructing other classes is called a \emph{metaclass}~\cite{pydatamodel}.
Metaclasses are a core concept of the Python type model and are strongly related to the changes that occured in the type/class unification process. 
Before Python 2.2, built-in types and user-defined classes operated under distinct object models. Built-ins were managed by the \texttt{type} system (the class \texttt{type} was the type of types), while user-defined classes had a separate internal construction mechanism. Afterward, \texttt{type} was siginificantly extended and generalized, becoming the sole metaclass responsible for all new-style classes. When Python 3 discontinued support for \texttt{classobj}, \texttt{type} became the supreme metaclass, which governed the creation of all classes.

Note: While there is the possibility to define a class as instance of another metaclass, this is a more advanced facility and is seldom used in practice. This was nicely noted by Tim Peters~\cite[p.~655]{ramalho2015fluent}:
\begin{quote}
\small
[Metaclasses] are deeper magic than 99\% of users should ever worry about. If you wonder whether you need them, you don't (the people who actually need them know with certainty that they need them, and don't need an explanation about why).
\end{quote}

\subsubsection{Duck Typing}
\label{sec:duck-typing}
Another core concept of Python is \emph{duck typing}~\cite{duckwiki}. In Python, the suitability of an object for a given operation is determined by its behavior (i.e. its methods and attributes), rather than its explicit class or inheritance hierarchy. This approach is encapsulated by the famous saying: \emph{if it walks like a duck and quacks like a duck, then it must be a duck}.

\begin{lstlisting}[language=Python, numbers=none, caption=Applied duck typing, label={ex:duck1}]
class Duck:
    def quack(self):
        return "Quack!"

class Donald:
    def quack(self):
        return "Nobody knows more about quacking than me."

def do_quack(x):
    print (x.quack())

duck = Duck()
do_quack(duck)  # OK
donald = Donald()
do_quack(donald)  # OK
\end{lstlisting}
The output for this snippet is:
\begin{lstlisting}[numbers=none]
Quack!
Nobody knows more about quacking than me.
\end{lstlisting}
Example~\ref{ex:duck1} illustrates the fact that the function \texttt{do\_quack} accepts both instances of \texttt{Duck} and \texttt{Donald}, since these classes both define the \texttt{quack} method with the required number of parameters. A question raised here: \emph{What is the type of }\texttt{x}\emph{? Does duck typing supply an answer for it?} 

\subsubsection{Type Annotations}
Type annotations, also known as \emph{type hints}, were introduced by Python Enhancement Proposal (PEP) 484~\cite{pep484}. This document defines a standard syntax for adding type hints to Python code, to simplify static type checking and improve code readability. As a result, errors can be detected ahead of time by static type checkers, while IDEs can provide better autocompletion and code refactoring capabilities. Needless to say, maintenance and collaborative development are also greatly improved by using type annotations. 
\begin{lstlisting}[language=Python, numbers=none, caption=Example of annotated code, label={ex:annotated1}]
def add(x: int, y: int) -> int:
    return x + y
\end{lstlisting}
In the above example, the code for function \texttt{add} is enriched with type annotations for the parameters and the return value. The type hints indicate that both parameters are expected to be integers and the function itself returns an integer value.

Note that type annotations are not enforced at runtime. They are purely optional and serve, at most, as a sort of documentation for the code that can be used by type checkers. This means that the following code will run just like the one in Example~\ref{ex:annotated1}, although there is no addition operation defined for dictionaries:
\begin{lstlisting}[language=Python, numbers=none, caption=Example of badly annotated code, label={ex:annotated2}]
def add(x: dict, y: dict) -> list:
    return x + y
\end{lstlisting}
\emph{Question: is it a common consensus that all type annotations represent Python types?}

\subsubsection{Abstract Base Classes}
\label{sec:abcs}
Abstract Base Classes (ABCs) provide mechanisms for explicitly defining interfaces or contracts that objects can adhere to, enhancing Python's type checking capabilities. ABCs were introduced by PEP 3119~\cite{pep3119} to formalize and standardize the inspection of object behavior. Prior to this, there were two ways to check whether an object adhered to a specific contract:
\begin{itemize}[leftmargin=*]
    \item manually inspecting the presence of certain methods. For example, checking whether the object's class defines the \texttt{\_\_len\_\_} method using Python's \texttt{hasattr}. This approach may become cumbersome in cases where multiple methods need to be checked;
    \item checking base classes throughout the inheritance tree. For example, verifying if the object's class is a direct or indirect subclass of \texttt{list}. This is undesirable because it brings along specific implementations and attributes that are beyond the scope of the contract.
\end{itemize}

ABCs also introduced \emph{virtual base classes}. This should not be confused with the C++ concept that bears the same name~\cite{virtualcpp}, as their similarity is only nominal. In Python, when a class \texttt{Foo} is registered as a virtual subclass by calling \texttt{FooABC.register(Foo)}, the ABC \texttt{FooABC} becomes the \emph{virtual base class} of \texttt{Foo}. As a result, \texttt{issubclass(Foo, FooABC)} will return \texttt{True}, even though \texttt{Foo} does not inherit from \texttt{FooABC}. This mechanism is facilitated by the metaclass \texttt{abc.ABCMeta}, of which all abstract base classes are instances. Notably, this occurs without requiring explicit inheritance.
\begin{lstlisting}[language=Python, numbers=none, caption=Different mechanisms for ABC recognition in class hierarchies, label={ex:abchooks}]
class MyABC(metaclass=ABCMeta):
    @abstractmethod
    def foo(self): ...

class MyABCHooked(metaclass=ABCMeta):
    @abstractmethod
    def foo(self): ...

    @classmethod
    def __subclasshook__(cls, subclass):
        if cls is MyABCHooked:
            if any("foo" in B.__dict__ for B in subclass.__mro__):
                return True
        return NotImplemented

class Sub1(MyABC):
    def foo(self):
        return 1

class Sub2:
    def foo(self):
        return 2

class Sub3:
    def foo(self):
        return 3

MyABC.register(Sub2)

print(f'Sub1 is subclass of MyABC: {issubclass(Sub1, MyABC)}')
print(f'Sub2 is subclass of MyABC: {issubclass(Sub2, MyABC)}')
print(f'Sub3 is subclass of MyABC: {issubclass(Sub3, MyABC)}')

print(f'Sub1 is subclass of MyABCHooked: {issubclass(Sub1, MyABCHooked)}')
print(f'Sub2 is subclass of MyABCHooked: {issubclass(Sub2, MyABCHooked)}')
print(f'Sub3 is subclass of MyABCHooked: {issubclass(Sub3, MyABCHooked)}')
\end{lstlisting}
This program outputs the following:
\begin{lstlisting}[language=, numbers=none]
Sub1 is subclass of MyABC: True
Sub2 is subclass of MyABC: True
Sub3 is subclass of MyABC: False
Sub1 is subclass of MyABCHooked: True
Sub2 is subclass of MyABCHooked: True
Sub3 is subclass of MyABCHooked: True
\end{lstlisting}
In Example~\ref{ex:abchooks}, \texttt{Sub1} explicitly inherits from the \texttt{MyABC} class. In contrast, \texttt{Sub2} does not explicitly inherit from it, but is registered as a virtual subclass by calling \texttt{MyABC.register(Sub2)}. This shifts the inheritance recognition responsibility to the ABC's side. The \texttt{\_\_subclasshook\_\_} method, demonstrated by \texttt{MyABCHooked}, requires no explicit inheritance on either side, as long as its logic returns \texttt{True} for a given \texttt{subclass}.

Note that the \texttt{register} and \texttt{\_\_subclasshook\_\_} mechanisms only affect runtime type checking using \texttt{issubclass}. Static type checkers, like Mypy~\cite{mypy}, do not use these methods for inheritance and static type checking. They generally require explicit inheritance and they have hardcoded knowledge for widely used data structures and ABCs. For example, Mypy knows that a \texttt{list} is a subclass of the \texttt{Collection} ABC defined in the \texttt{collections.abc} Python module, even though \texttt{list} does not inherit from it directly.

\subsubsection{Protocols}
\label{sec:protocols}
Protocols~\cite{pep544} are the conceptual descendants of ABCs. They are built upon the ABC infrastructure, but are semantically different. They were created to aid type checkers, with the primary objective of formalizing duck typing for static analysis. Therefore, Protocols represent a shift from inheritance-based typing toward structural typing, where adherence to an interface is determined by the presence and type compatibility of certain members. This provides stronger guarantees of type safety before runtime, because conformance to expected interfaces can be validated even when no explicit inheritance is declared. As a consequence, Python developers can leverage the benefits of duck typing with the added safety and predictability of static analysis.

By default, using protocols as arguments for \texttt{issubclass} or \texttt{isinstance} raises a runtime error. However, they can be decorated with \texttt{@runtime\_checkable}, which enables runtime checking. This functionality is supported by the \texttt{\_\_subclasshook\_\_} mechanism within the metaclass \texttt{\_ProtocolMeta}, itself a subclass of \texttt{ABCMeta}. Note that, although the same runtime checking mechanism is used, the most significant contribution of protocols is that they provide a formal mechanism for structural subtyping, greatly enhancing the precision of static type checking.

\begin{lstlisting}[language=Python, numbers=none, caption=Using protocols for runtime and static type checking, label={ex:protocols}]
@runtime_checkable
class MyProtocol(Protocol):
    def foo(self, x: int) -> bool: ...

class Sub1:
    def foo(self, x: float) -> int:
        return 1

class Sub2:
    def foo(self, x: str) -> int:
        return 2

class Sub3:
    def foo(self, x: int) -> bool:
        return True

def f1(x: MyProtocol):
    return None

f1(Sub3())
f1(Sub2())
print(f'Sub1 is subclass of MyProtocol: {issubclass(Sub1, MyProtocol)}')
print(f'Sub2 is subclass of MyProtocol: {issubclass(Sub2, MyProtocol)}')
print(f'Sub3 is subclass of MyProtocol: {issubclass(Sub3, MyProtocol)}')
\end{lstlisting}
The above example enhances the one in Example~\ref{ex:abchooks}. In this case, we used a decorated protocol which describes the expected structure of the parameters for the \texttt{f1} function. At runtime, all the subclass checks return \texttt{True} because they only check for the presence of the required method name, not its full type signature. However, Mypy is not so lenient; running it to statically check this code outputs the following:
\begin{lstlisting}[language=]
error: Argument 1 to "f1" has incompatible type "Sub2"; expected "MyProtocol"  [arg-type]
note: Following member(s) of "Sub2" have conflicts:
note:     Expected:
note:         def foo(self, x: int) -> bool
note:     Got:
note:         def foo(self, x: str) -> int
Found 1 error in 1 file (checked 1 source file)
\end{lstlisting}
Thus, Mypy correctly detects that \texttt{Sub2} does not conform to the protocol due to incompatible argument type annotations.

\emph{Question: The intertwining of runtime and static mechanisms for interface conformance raises a deeper question about Python's type system: what, ultimately, constitutes a type in Python?}

\subsection{Types and Classes in Python}
The concept of \emph{type} in Python, particularly with the introduction of ABCs and Protocols, extends beyond a simple mapping to a class. It is crucial to understand the differences between Python's runtime type system, the role played by classes and the purpose of static type annotations.

\subsubsection{Classes as Types}
In Python, classes serve as blueprints for both runtime behavior and, as explained earlier, static type checking. However, the notion of \emph{type} remains context-dependent.

\paragraph{Runtime types.} The runtime type of an object is retrieved by calling the built-in \texttt{type} function, which returns the class of that specific object. For example:
\begin{itemize}[leftmargin=*]
    \item \texttt{type(5)} returns \texttt{<class 'int'>};
    \item \texttt{type('xyz')} returns \texttt{<class 'str'>}.
\end{itemize}
As previously discussed, classes are objects themselves. They are instances of a metaclass. Therefore, when calling \texttt{type} with a class as an argument, its metaclass will be returned:
\begin{itemize}[leftmargin=*]
    \item \texttt{type(int)} returns \texttt{<class 'type'>};
    \item \texttt{type(str)} returns \texttt{<class 'type'>} as well.
\end{itemize}
For most built-in and user-defined classes, the metaclass is \texttt{<class 'type'>}, which is a direct result of the \emph{type/class unification process} described in Section~\ref{subsec:history}.

In conclusion, we might consider that, at runtime, \emph{the type of an object is the class from which the object was instantiated}. However, the relationship is more intricate when considering the foundational classes:
\begin{itemize}[leftmargin=*]
    \item \texttt{object} is the base class of every Python class, including metaclasses, and thus \texttt{<class 'type'>} itself;
    \item \texttt{object} is an instance of \texttt{<class 'type'>}, just like most built-in classes.
\end{itemize} 
This leads to a fundamental question: \emph{what is }\texttt{<class 'type'>}\emph{ an instance of}? The answer is as simple as it is surprising: \texttt{<class 'type'>} \emph{is an instance of itself}. This is a special case hardcoded into the Python interpreter to support the unified object model.
\begin{lstlisting}[language=Python, numbers=none, caption={Runtime instance and subclass checks}]
>>> type(int) --> <class 'type'>
>>> type(type) --> <class 'type'>
>>> type(object) --> <class 'type'>
>>> issubclass(type, object) --> True
\end{lstlisting}

Conceptually, a class defines a type in Python. For example, the values of Python's \texttt{int} class can be formally expressed using its constructor, which creates objects (with specific attributes and methods) corresponding to mathematical integers: 
\[
    Val(\mathtt{int}) = \{ \mathtt{int}(x) \mid x \in \mathbb{Z} \}
\]
The values of a user-defined class can be described in a similar fashion:
\begin{lstlisting}[language=Python, numbers=none, belowskip=-1.5\baselineskip, label={ex:point}] 
class Point:
    def __init__(self, x: int, y: int):
        self.x = x
        self.y = y
\end{lstlisting}
The values of the \texttt{Point} type defined above are all instances of class \texttt{Point} where its attributes are values of the \texttt{int} class: \[
    Val(\mathtt{Point}) = \{ \mathtt{Point}(x, y) \mid x, y \in Val(\mathtt{int}) \}
\]

\paragraph{Not-so-runtime types.} Type annotations add more complexity to the definition of what constitutes a type in Python. For example, \texttt{Sized} is not a type that can be instantiated at runtime. It is an ABC, which implements the \texttt{\_\_subclasshook\_\_} mechanism. It describes that an object has a length, or size, that can be retrieved by calling the built-in \texttt{len} function. And yet, it is considered acceptable to use it to describe the type of a function parameter:
\begin{lstlisting}[language=Python, numbers=none, caption=Using an ABC as a type hint]
def foo(x: Sized) -> int:
    return len(x)
\end{lstlisting}
Naturally, classes that can be instantiated, such as \texttt{int}, \texttt{float}, \texttt{list} and the like, may also be used as type annotations. But, should protocols or ABCs not be considered types? This would severely limit our possibilities. There are numerous type hierarchies that include protocols amongst others. For example, \texttt{list} can be seen as a subtype~\cite{liskov1994} of \texttt{Sized}, because the property that they have a measurable length is true for \texttt{list} values. Therefore, a value of type \texttt{list} is a valid substitute for any \texttt{Sized} type requirement.

\begin{figure}[ht!]
    \centering
    \includegraphics[scale=.35]{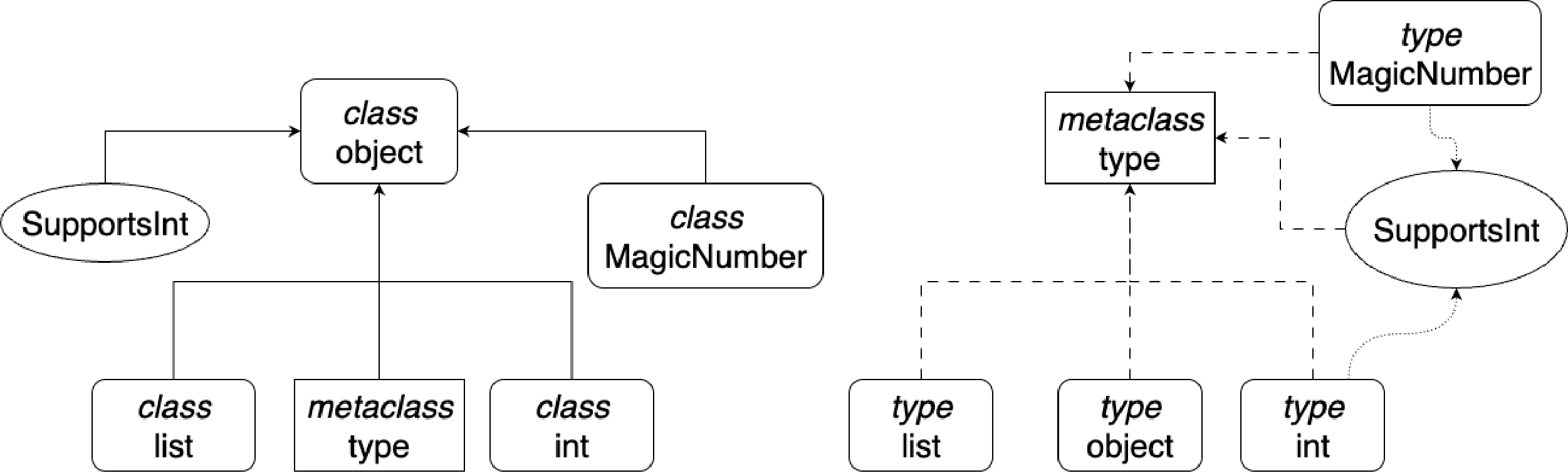}
    \caption{Example of different relations between classes and types}
    \label{fig:bush}
\end{figure}

\paragraph{Bushwhacking through the jungle.} To make sense of the intricate landscape of Python types, we identify a few key relational paths through the jungle. Specifically, we distinguish three kinds of relationships between classes and types:
\begin{itemize}[leftmargin=*]
    \item \emph{subclass-of}, which indicates that a class explicitly inherits another (for example, every class inherits from \texttt{object});
    \item \emph{object-instance-of}, which indicates that a class is an instance of another (for example, every class is an instance of the metaclass \texttt{type});
    \item \emph{type-instance-of}, which indicates that a class adheres to the structural contract of another (for example, \texttt{list} conforms to the \texttt{Sized} ABC by defining the \texttt{\_\_len\_\_} method).
\end{itemize}
These relations are exemplified in Figure~\ref{fig:bush}. In this figure, rectangles represent runtime classes and types (with the metaclass rectangle distinguished by sharper corners), while ovals represent uninstantiable classes, such as the \texttt{SupportsInt} protocol. Solid arrows represent a \emph{subclass-of} relation, dashed arrows describe an \emph{object-instance-of} relation, and dotted arrows are \emph{type-instance-of}.
This distinction helps clarify the dual nature of protocols and ABCs. They can be interpreted either as types, via structural or virtual conformance, or as classes, via inheritance.
For instance, consider a class \texttt{MagicNumber} that defines an \texttt{\_\_int\_\_} method, but does not inherit from \texttt{SupportsInt}:
\begin{lstlisting}[language=Python, numbers=none]
class MagicNumber:
    def __init__(self, nr: int):
        self.remaining = nr

    def __int__(self):
        return self.remaining

def get_horcrux_nr(x: SupportsInt) -> str:
    return (f'There are {x} horcruxes remaining')

foo = MagicNumber(4)
print(get_horcrux_nr(foo))  # prints 'There are 4 horcruxes remaining'
\end{lstlisting}
Despite not being a subclass of \texttt{SupportsInt}, this class is still accepted by the function \texttt{get\_horcrux\_nr} by both runtime and static checks. This is because it satisfies the \texttt{SupportsInt} structurally. This is precisely what Figure~\ref{fig:bush} captures: even though the class \texttt{MagicNumber} is not connected to \texttt{SupportsInt} via \emph{subclass-of}, it is linked to it as a type, highlighting the dual nature of protocols: they can act both as classes and as type specifications. It also illustrates how classes, more broadly, can be regarded both as runtime constructs and as types within the system.

\section{Toward a Formal Python Type System}
The previous section charted the landscape of Python's type mechanisms, from runtime constructs such as classes and metaclasses, to static ones like protocols. In this section, we take a step toward building the foundation of what a \emph{type} means in Python by introducing a formal framework rooted in established type-theoretic ideas. Using abstract data types (ADTs) and their representation as existential types, we present a formalization that models static Python types in a way that is faithful to Python's design. Our focus is restricted to programs where types are not dynamically altered at runtime, thereby ensuring that the formalization applies to a well-defined and analyzable fragment of the language.

\subsection{Foundational Concepts}
\label{subsec:adtfoundation}
A suitable formal system for Python types must be found at the intersection of the following concepts:
\begin{enumerate}
    \item \emph{Every type in Python is represented by a class}.
    \item \emph{A class is an implementation of an abstract data type}.
    \item \emph{An Abstract Data Type (ADT) defines a data type by its behavior rather than its implementation}. 
    \item  \emph{A formal type system classifies entities into types, where each type represents a collection of values and a set of operations that can be performed on them.}
\end{enumerate}
The relationship between ADTs and type systems is best expressed by \emph{abstract types have existential type}, a theory developed by John C. Mitchell and Gordon D. Plotkin~\cite{mitchellexistential} (see also~\cite{pierce-tpl}). Since we are interested in a static type system for Python, we are focusing only on the static aspect of the ADT, ignoring the specification of the behavior. For instance, if an ADT $t$ with the operations $x_1,\ldots, x_n$ is described as an expression of the form~\cite{mitchellexistential}\\
\[
    \textbf{abstype } t \textbf{ with } x_1: \sigma_1, \ldots, x_n: \sigma_n \textbf{ is }M\textbf{ in }N
\]
where $M\textbf{ in }N$ is a data algebra expression specifying the behavior, then
$\textbf{abstype } t \textbf{ with } x_1: \sigma_1, \ldots, x_n: \sigma_n$
describes only the static aspect of the ADT. It is rather a type specification for the ADT.

\paragraph{Existential Types.} An \emph{existential type}, denoted as $\exists X. \tau$, can be read as "there exists a type $X$ such that $\tau$ holds". In the context of Python, this translates to:
\begin{itemize}
\item $\exists X$: there exists some concrete data type $X$ chosen as the \emph{representation type}. This is the hidden implementation of our ADT. In Python, for example, a class can serve as such a representation;
\item $\tau$: describes the signature of the ADT's operations. Externally, these operations are typed in terms of the abstract type. Internally, they are implemented using the representation type $X$.
\end{itemize}

The intuition for an element of $\exists X. \tau$ is a pair $(S, t)$, consisting of:
\begin{itemize}
    \item a concrete type $S$, which substitutes the abstract type $X$;
    \item a term $t$ of type $\tau[S/X]$, which represents the concrete implementation of the type, where every free occurrence of $X$ is substituted by $S$.
\end{itemize}
\setcounter{example}{\value{lstlisting}}
\begin{example}
\label{ex:existential_duck}
The classes \texttt{Duck} and \texttt{Donald} from \Cref{sec:duck-typing} represent the same existential type:
\[
    \mathtt{QuackET} = \exists Q.\{\mathtt{quack}: Q \to \mathtt{StrET}\}
\]
Here, \texttt{StrET} denotes the existential type of Python string values. We consider that the Python class \texttt{Duck} implements a concrete type with the same name, which is an element of the existential type \texttt{QuackET}: 
\begin{itemize}[leftmargin=*]
    \item We choose the concrete type $S = \mathtt{Duck}$. In Python code, \texttt{Duck} is a class definition. Here, however, \texttt{Duck} plays the role of the representation type $S$ that witnesses the existential type. This is intentional, since Python classes actually define types~\cite{pyclasses}.
    
    \item The term $t$ must have the type $\tau[\mathtt{Duck}/Q]$, i.e. $\tau[\mathtt{Duck}/Q] = \{ \mathtt{quack}: \mathtt{Duck} \rightarrow \mathtt{StrET} \}$. This term is a record type that maps \texttt{quack} to its implementation. In this case, the implementation is the \texttt{quack} method from the \texttt{Duck} class: $t = \{ \mathtt{quack} := \mathtt{Duck.quack} \}$. This should be read as a binding of the operation symbol \texttt{quack} to its implementation in the \texttt{Duck} class.
    
    \item Therefore, an element of \texttt{QuackET} is the pair $
    (\mathtt{Duck}, \{ \mathtt{quack} := \mathtt{Duck.quack} \})$.
\end{itemize}
\end{example}

\paragraph{Generic Existential Types.} If $\exists X.\tau$ is an existential type and $\tau$ includes free type variables of the form $Y_1, \ldots, Y_k$ that are distinct from $X$, then $\forall Y_1, \ldots, Y_k. \exists X. \tau$ is a \emph{generic existential type}:
\begin{itemize}[leftmargin=*]
    \item $\forall Y_1, \ldots, Y_k$: means \emph{for all types $Y_1, \ldots, Y_k$}, where $Y_i$ are generic type parameters;
    \item $\exists X$: translates to \emph{there exists a hidden implementation type $X$}, whose own structure depends on the $Y_i$ type parameters;
    \item $\tau$ is the public interface whose operations are defined in terms of both the public types $Y_i$ and the hidden type $X$.
\end{itemize}

Conceptually, universal quantification ($\forall$) introduces parametric polymorphism, while existential quantification ($\exists$) hides the representation type.

The intuition for an element of $\forall Y.\tau'$ is a function that, given a type $Z$, produces a concrete instance of $\tau'[Z/Y]$. It follows that the intuition for an element of $\forall Y.\exists X.\tau$ is a function that, for each type $Z$, produces a pair consisting of a type $S$ and a term $t$ of type $\tau[Z/Y][S/X]$.
\begin{example}
\label{ex:supportsabs}
In Python, \texttt{SupportsAbs} is a generic protocol with one parameter, where the type variable is used to denote the return value type of the \texttt{\_\_abs\_\_} method. A possible existential type for it is:
\[
\mathtt{SupportsAbsET} = \forall Y. \exists X. \{ \mathtt{\_\_abs\_\_}: X \rightarrow Y \}
\]
The existential type of all instances whose absolute values are \texttt{float} objects is obtained by instantiating the type parameter $Y$ with $\mathtt{FloatET}$:
\[
    \mathtt{SupportsAbs[FloatET]} = \exists X.\{ \mathtt{\_\_abs\_\_}: X \rightarrow \mathtt{FloatET} \}
\]
Naturally, \texttt{FloatET} represents the existential type of Python \texttt{float} values.
The built-in Python classes \texttt{float} and \texttt{complex} act as the representation types that serve as witnesses of this existential type, using the same logic as in Example~\ref{ex:existential_duck}.
\end{example}

In order to model inheritance, we are using the bounded quantification $\exists X <: T.\tau$~\cite{pierce-tpl,cardelli}. We read this as: even if $X$ is abstract, we know that it is a subtype\footnote{Due to the space limit, the definition for subtyping is not included, but it follows the lines of~\cite{cardelli} and~\cite{cardellisemantics}.} of $T$.

\subsection{A Proposal for a (Static) Python Type System}
\label{sec:proposal}
This subsection advances a proposal for a static type system for Python, called \emph{Pythonic Type System (PyTS)}. Our goal is to capture the subset of Python types that can be modeled using \emph{existential types}. 

\subsubsection{Built-in Existential Types}
A formal signature $\tau$ of an ADT is a record type that maps class member names to type expressions that are built from a set of fundamental constructs. In Python, every type is defined by a class. This applies to all types, including the built-in ones~\cite{pydatamodel}, which are classes implemented in the CPython~\cite{cpythongit} backend.
For example, \texttt{int} is a built-in Python class and its objects are stored as C structures in the backend. We consider that \texttt{int} implements an existential type as follows:
\[
    \mathtt{IntET} = \exists IT.\{ \mathtt{\_\_repr\_\_}: IT \rightarrow \mathtt{StrET}, \ldots \}
\]
Naturally, \texttt{IntET} has many other members. We chose to describe this specific method to highlight the fact that some members may depend on other existential types. So, \texttt{\_\_repr\_\_} outputs a string value, which is an instance of Python's \texttt{str} class. The \texttt{str} class can be viewed as implementing an existential type as well, which we denoted \texttt{StrET}:
\[
    \mathtt{StrET} = \exists S.\{ \mathtt{\_\_len\_\_}: S \rightarrow \mathtt{IntET}, \ldots \}
\]

We observe that \texttt{StrET} and \texttt{IntET} are mutually defined. We propose a type system, \emph{Pythonic Type System (PyTS)}, which uses Python-specific type expressions to build existential types. These expressions are built using a set of primitives derived from Python's core classes~\cite{pydatamodel}, excluding the \texttt{type} metaclass. The primitives themselves are also existential types, and are divided into the following categories:
\begin{itemize}[leftmargin=*]
    \item \textbf{Built-in Atomic Existential Types} represent fundamental types:
    \begin{itemize}[leftmargin=*]
         \item \emph{numeric types}: \texttt{BoolET}, \texttt{IntET}, \texttt{FloatET}, \texttt{ComplexET};

        \item \emph{scalar sequence types}: \texttt{StrET}, \texttt{BytesET};

        \item \texttt{ObjectET}, for the \texttt{object} class, which is the base class for all Python classes;

        \item \texttt{BottomET}, with the defining characteristic is that it contains no values. In Python type annotations, \texttt{typing.Never} denotes the bottom type for static type checkers;

        \item \texttt{NoneTypeET}, corresponding to Python's \texttt{NoneType} class. This class has a single possible value, \texttt{None}, and represents the absence of a meaningful result and fulfills the role of the unit type in our system.
    \end{itemize}
    
    \item \textbf{Built-in Generic Container Existential Types} act like existential type constructors that can be parameterized by other existential types:
    \begin{itemize}[leftmargin=*]
        \item \emph{sequence types}: \texttt{ListET}, \texttt{TupleET}, \texttt{BytearrayET};

        \item \emph{set types}: \texttt{SetET}, \texttt{FrozensetET};

        \item \emph{mapping type}: \texttt{DictET}.
    \end{itemize}
\end{itemize}

We used the following naming convention to name these types: each type name is capitalized and formed by taking the corresponding Python class name and appending the suffix \texttt{ET}.

A generic type constructor can be specialized to form a concrete type expression. For example:
\begin{itemize}[leftmargin=*]
    \item[] \texttt{ListET[IntET]} constructs the existential type for instances of lists of integers.

    \item[] \texttt{TupleET[IntET, StrET]} constructs the existential type for tuples that contain an integer on the first position and a string on the second.
\end{itemize}
\paragraph{Remarks}
\begin{enumerate}[leftmargin=*]
    \item We use an ellipsis ($\ldots$) in two ways:
    \begin{itemize}[leftmargin=*]
        \item inside the \emph{parameter list} to describe existential types for tuples which hold an unknown number of elements of a certain type. For example, $\mathtt{TupleET[IntET, \ldots]}$ is the existential type for tuples that contain an arbitrary number (including zero) of integer values;
        \item inside \emph{type signatures} to indicate that additional members are present, but omitted for brevity. For example, $\mathtt{IntET} = \exists IT.\{ \mathtt{\_\_repr\_\_}: IT \rightarrow \mathtt{StrET}, \ldots \}$ specifies only one method of the full signature, leaving the others implicit.
    \end{itemize}

    \item We use \texttt{ObjectET} as the top type in our type system because \texttt{object} is the base class of every Python class.
    In Python static type checking, \texttt{Any} acts as a universal wildcard, permitted as an annotation anywhere a type is expected. It is used to describe names or expressions whose types are not known statically, thereby aiding the gradual typing of Python programs~\cite{typingconcepts}. 
\end{enumerate}

The \texttt{PyTS} framework provides a layered model for Python's type system by distinguishing between the \emph{types of data values} (like \texttt{5}, \texttt{[1, 2, 3]}, or \texttt{Point(5, 10)}), and the \emph{types of class objects} that create them (like the \texttt{int}, \texttt{list}, or \texttt{Point} classes).  

\subsubsection{The Foundational (Blueprint) Layer}
\label{sec:blueprint}
This layer provides the formal types, or blueprints, for data values. The signatures of these existential types are record types containing type expressions, which are constructed from the following:
\begin{itemize}[leftmargin=*]
    \item \emph{Built-in Existential Types}: these are the fundamental existential types described above;

    \item \emph{Product Types}: these are types formed by the Cartesian product of two or more type expressions, for example $A \times B$. A value of this type contains a value from each constituent type in a specific, ordered sequence. We mainly use this construct to model domains for functions that accept multiple arguments;
    
    \item \emph{Sum Types}: these types, denoted by $A + B$, represent a disjoint union of other type expressions. A value of a sum type holds a value from either type $A$ or $B$. The closest correspondent to sum types in Python is \texttt{types.UnionType}~\cite{pep604}\cite{pystdunions}, primarily intended for type annotations;
    
    \item \emph{Function Types}: these types, denoted by $A \rightarrow B$, represent a mapping from an input type $A$ (the domain) to an output type $B$ (the codomain). In Python, this concept is described for type annotations using \texttt{typing.Callable}. For example, a function taking an integer and returning a string would be annotated as \texttt{Callable[[int], str]}. Note: for functions that accept a variable number of arguments, as is common in Python with \texttt{*args} and \texttt{**kwargs}, the signature $\mathtt{TupleET[ObjectET, \ldots]} \times \mathtt{DictET[StrET, ObjectET]}$ describes the arbitrary positional and keyword arguments;

\end{itemize}
Using the constructs described above, the existential type \texttt{ObjectET} is described as follows:
\begin{align*}
    &\mathtt{ObjectET} = \exists O.\{\\
    &\qquad \mathtt{\_\_new\_\_}: \mathtt{TupleET[ObjectET, \ldots]} \times \mathtt{DictET[StrET, ObjectET]} \rightarrow O, \\
    &\qquad \mathtt{\_\_init\_\_}: O \times \mathtt{TupleET[ObjectET, \ldots]} \times \mathtt{DictET[StrET, ObjectET]} \rightarrow O, \ldots \}
\end{align*}

Using the constructs described above, $\mathtt{PyTS}$ introduces the following fundamental existential types:
\begin{itemize}[leftmargin=*]
    \item \texttt{TypeVarET}, which is the existential type of \texttt{TypeVar} values. These are used to describe the generics used for static type checking:
    \[
        \mathtt{TypeVarET} = \exists TV <: \mathtt{ObjectET}.\{ \mathtt{\_\_name\_\_}: \mathtt{NoneTypeET} \rightarrow \mathtt{StrET}, \ldots \}
    \]

    \item \emph{The Generic Existential Type Family}, denoted $\mathtt{GenericET}$, consists of the types $\mathtt{GenericET_n}$ for all natural numbers $\mathtt{n} \geq 1$. Each $\mathtt{GenericET_n}$ denotes the existential type of generic classes with exactly \texttt{n} parameters:
    \begin{align*}
        &\mathtt{GenericET_n} = \forall T_1,\ldots,T_n.\exists G <: \mathtt{ObjectET}.\{\\
        &\qquad \mathtt{\_\_parameters\_\_}: \mathtt{NoneTypeET} \rightarrow \mathtt{TupleET_n[TypeVarET, \ldots]}, \ldots \}
    \end{align*}
    where $\mathtt{TupleET_n}$ denotes a tuple with the arity \texttt{n}. We treat generics as having at least one parameter. The odd fact that, in Python, \texttt{typing.Generic} itself is instantiable without parameters is a runtime quirk that we do not model in our type system. Also, it is seldom the case for instantiating \texttt{typing.Generic}. It is rather used as a parent class for other generic classes and subclassing it without at least one type parameter is not permitted. This rule is bypassed only for specific cases, such as \texttt{typing.Protocol};

    \item \emph{The Protocol Existential Type Family}, denoted \texttt{ProtocolET}, consists of the types $\mathtt{ProtocolET_n}$, for all natural numbers $\mathtt{n} \geq 0$. For generic protocols with exactly \texttt{n} parameters, the existential type is: 
    \begin{align*}
        &\mathtt{ProtocolET_n} = \forall T_1, \ldots, T_n. \exists P <: \mathtt{GenericET_{n+1}[P, \mathit{T_1}, \ldots, \mathit{T_n}]}.\{\\
        &\qquad \mathtt{\_\_new\_\_}: P \times \mathtt{TupleET[ObjectET, \ldots]} \times \mathtt{DictET[StrET, ObjectET]} \rightarrow \mathtt{BottomET},\\
        &\qquad \mathtt{\_is\_protocol}: \mathtt{NoneTypeET} \rightarrow \mathtt{BoolET},\\
        &\qquad \mathtt{\_is\_runtime\_protocol}: \mathtt{NoneTypeET} \rightarrow \mathtt{Bool}, \ldots \}
    \end{align*}
    \paragraph{Remarks.}
    \begin{itemize}[leftmargin=*]
        \item In our type system, whenever we explicitly mention a member of the child class that is also present in the parent class, means the inherited one is overriden. In this example, we used the return type \texttt{BottomET} for the \texttt{\_\_new\_\_} method to describe that Python protocols are not instantiable;
        \item For $\mathtt{n}=0$, the type $\mathtt{ProtocolET_0}$ is not generic.
    \end{itemize}
\end{itemize}
\begin{example}
\label{ex:mylistclass}
The following Python class:
\begin{lstlisting}[language=Python, numbers=none, belowskip=-1.5\baselineskip]
class MyList(list):
    pretty_string = lambda self: "test"
\end{lstlisting}
produces values of the existential type:
\[
\mathtt{MyListET} = \forall T.\exists L <: \mathtt{ListET[\mathit{T}]}.\{ \mathtt{pretty\_string}: M \rightarrow \mathtt{StrET}, \ldots \}
\]
The class \texttt{MyList} inherits from Python's \texttt{list} class, whose values are modeled by the generic existential type \texttt{ListET}. Because \texttt{ListET} is parameterized by a type variable $T$, the resulting \texttt{MyListET} type is generic as well. 
\end{example}
\begin{example}
Consider the following Python protocols:
\begin{lstlisting}[language=Python, numbers=none, belowskip=-1.5\baselineskip]
class SupportsInt(Protocol):
    def __int__(self) -> int: ...

class SupportsAbs(Protocol[T]):
    def __abs__(self) -> T: ...
\end{lstlisting}
We model the existential type for \texttt{SupportsInt} as:
\[
    \mathtt{SupportsIntET} = \exists SI <: \mathtt{ProtocolET_0}.\{ \mathtt{\_\_int\_\_}: SI \rightarrow \mathtt{IntET}, \ldots \}
\]
For \texttt{SupportsAbs}, which is generic in the return type of the \texttt{\_\_abs\_\_} method, we provide a generic existential type with one parameter:
\[
    \mathtt{SupportsAbsET} = \forall T.\exists SA <: \mathtt{ProtocolET_1[T]}. \{ \mathtt{\_\_abs\_\_}: SA \rightarrow T, \ldots \}
\]
Using $\mathtt{PyTS}$, we are able to provide a more accurate existential type than the one we proposed in Example~\ref{ex:supportsabs}.
\end{example}

\subsubsection{The Meta Layer}
The previous layer provides a formal type system for "ordinary" class instances in Python. However, since \emph{everything is an object} in Python, so are classes themselves. We write \emph{class-as-value} whenever we refer to a class as a \emph{first-class object}.
\begin{example}~
\label{ex:classobj}
\begin{lstlisting}[language=Python, numbers=none, belowskip=-1.5\baselineskip]
class MyList(list):
    pretty_string = lambda self: "test"

foo = MyList([1, 2, 3])  # MyListET[IntET]
bar = [MyList]  # what is the type of the class-as-value stored in bar?
\end{lstlisting}
\end{example}
To answer the question in the above example, we must consider the \emph{dual role} that a class plays in Python:
\begin{enumerate}[leftmargin=*]
    \item \emph{the blueprint role}: first, a class serves as a blueprint that defines the structure and behavior of its instances. This is the role captured by the first layer of our type system. In the example above, the variable \texttt{foo} holds a data value that conforms to the \texttt{MyList} blueprint.
    \item \emph{the value role}: second, a class itself is a value that exists at runtime, that can be manipulated and can perform actions.
\end{enumerate}
In Python, all classes are instances of the \texttt{type} metaclass.
The \texttt{class} statement is equivalent to invoking the metaclass (by default, \texttt{type}) with arguments corresponding to the following parameters:
\begin{itemize}[leftmargin=*] 
    \item \emph{name}: a string parameter, for the class name;
    \item \emph{bases}: a tuple of base classes from which to inherit. Left empty, the \texttt{object} class is added by default;
    \item \emph{dict}: a dictionary that contains attribute and method definitions for the class body;
    \item \emph{**kwargs} is an optional parameter which represents a list of keyword-only arguments that are passed to the appropriate metaclass machinery.
\end{itemize}
A rewrite of Example~\ref{ex:classobj} is:
\begin{example}~
\label{ex:classtypeobj}
\begin{lstlisting}[language=Python, numbers=none, belowskip=-1.5\baselineskip]
MyList = type('MyList', (list, ), {'pretty_string': lambda self: "test"})

foo = MyList([1, 2, 3])  # MyListET[IntET]
bar = [MyList]  # what is the type of the class-as-value stored in bar?
\end{lstlisting}
\end{example}
Note that calling $\mathtt{type(\ldots)}$ directly invokes the built-in metaclass \texttt{type} to create a new class-as-value.
In $\mathtt{PyTS}$, we add the following existential type corresponding to the metaclass type:
\begin{align*}
    &\mathtt{TypeET} = \exists M <: \mathtt{ObjectET}.\{ \mathtt{\_\_new\_\_}: \mathtt{StrET} \times \mathtt{TupleET[TypeET, \ldots]} \times \\
    &\qquad \mathtt{DictET[StrET, ObjectET]} \times \mathtt{DictET[StrET, ObjectET]} \rightarrow M, \ldots\}\texttt{ is }\mathtt{PyTS}
\end{align*}
where the annotation $\texttt{is }\mathtt{PyTS}$ means that the elements of $M$ are representations of classes-as-values.
\paragraph{Remarks.}
\begin{itemize}[leftmargin=*]
    \item \texttt{type} is the \emph{canonical witness} of this existential type. Other witnesses include its subtypes, such as \texttt{ABCMeta}, \texttt{\_ProtocolMeta} or even user-defined metaclasses;
    \item \texttt{type} is both a value of \texttt{TypeET} and the mechanism for creating other values of \texttt{TypeET};
    \item the second parameter, \emph{bases}, supplies the base classes given as a tuple of classes-as-values. Therefore, its type is $\mathtt{TupleET[TypeET, \ldots]}$.
\end{itemize}
In Example~\ref{ex:classtypeobj}, the class \texttt{MyList} is a value of \texttt{TypeET} and has the corresponding \texttt{MyListET} blueprint in $\mathtt{PyTS}$.
The answer to the question posed in this example: the type of values held by \texttt{bar} is \texttt{ListET[TypeET]}.

\section{Related Work}
\subsection{Static Type Checkers: Mypy and Pyright}
The most powerful driving force behind the evolution of Python's typing evolution are \emph{static type checkers}. These are used to parse Python code, especially annotated one, and report type inconsistencies before execution. Among them, Mypy~\cite{mypy} and Pyright~\cite{pyright} stand out.

\emph{Mypy.} Jukka Lehtosalo started development on this project started out in 2012. Later on, Guido van Rossum contributed by making Mypy the original static type checker for Python. Its primary function is that of a type checker, not a type inferencer, relying heavily on the annotations provided by developers. Mypy's implementation demonstrated the challenge this paper addresses: to correctly model Python's type system, Mypy needs to encode vast amount of knowledge about the behavior of built-in types and standard library modules. As we already presented in Section~\ref{sec:abcs}, it has hardcoded knowledge for widely used ABCs where runtime checks would fail for static analysis. Having to treat these special cases means that a cohesive, formal foundation could only improve future static analysis implementations.

\emph{Pyright.} This is a more recent static type checker, developed and maintained by Microsoft. It has become widely adopted and it is known for its high performance and its role as the engine behind the primary Python language server for Visual Studio Code, Pylance~\cite{pylance}. Like Mypy, Pylance is fundamentally a type checker that validates code against provided type annotations, conforminf to the standards set by the official PEPs.

\subsection{Formal and Theoretical Work}
To establish a formal foundation for the Python type system, we use insights from well-known literature on type theory. Although much of this work significantlly predates Python, it provides the necessary conceptual tools that provide a theoretical backdrop for our effort in reconciling Python's flexible typing and a more formal structure.

\emph{Abstract Types Have Existential Type}, by John C. Mitchell and 
Gordon D. Plotkin~\cite{mitchellexistential}, throughly formalizes the connection between abstract data types and existential qualification. It uses this formalization to explain information hiding in typed programming languages. This paper defines an ADT implementation as a \emph{data algebra}, which is a set of values and a set of operations that act upon these values. The main idea is that these data algebras can be assigned \emph{existential types}, that convey how the operations can be used without revealing the concrete representation type. As mentioned in Section~\ref{subsec:adtfoundation}, this paper is heavily inspired by these concepts.

\emph{On Understanding Types, Data Abstraction, and Polymorphism}, by Luca Cardelli and Peter Wegner~\cite{cardelli}, provides a comprehensive taxonomy of polymorphism in programming languages, distinguishing between subtype polymorphism (e.g. via inheritance and interface conformance) and parametric polymorphism (e.g. in generic functions and containers). For this purpose, the authors introduce \emph{Fun}, a $\lambda$-calculus-based engine that includes abstract data types, parametric polymorphism, and multiple inheritance. It also discusses type checking and inference mechanisms and shows how Fun can be used to model features from other programming languages, like ML, Ada or Simula. As Python combines both parametric and subtype polymorphism in a dynamic setting, these concepts and formal tools are directly relevant to our work.

\emph{A Behavioral Notion of Subtyping}, by Barbara H. Liskov and Jeannette M. Wing~\cite{liskov1994}, argues that the subtype relation between two types is simply a question of semantics. The core idea is that objects of a subtype should behave indistinguishably from those of their supertype, from the perspective of an user who interacts with supertype object. The paper highlights that traditional subtyping rules, with focus on method signatures, are not enough because they only prevent typing errors, not correct program behavior. The authors propose a framework where a subtype must preserve the behavioral specifications of its supertype, respecting contracts expressed via preconditions, postconditions, and invariants. While Python, nor this paper's formalisms, enforce behavioral subtyping, this work is nevertheless crucial for any formalization that aspires for semantic soundness. 

\emph{Static Type Analysis by Abstract Interpretation of Python Programs}, by Rapha{\"e}l Monat~\cite{monatecoop}, is a research paper that views type analysis as an instance of abstract interpretation, concentrating on detecting uncaught exceptions and proving program properties. While it does define a concrete semantics for a large subset of Python, which serves as the foundation for analyses, it does not aim to formalize the Python type system itself. A standout feature of this work is its aim to perform automatic analysis without requiring any type annotations using the Mopsa framework~\cite{mopsa}. This aligns with our goal of creating a type inference framework, although Monat's work clearly states analyzing functions in isolation is not an objective.

\section{Conclusion and Future Work}
This paper set out to address the theoretical foundation of Python's typing system. We first provided a brief history of the type system, while also explaining key typing concepts such as \emph{metaclasses}, \emph{duck typing}, \emph{abstract base classes}, and \emph{protocols}. Afterward, we set out to understand classes, how they define types and how are values seen by Python. Additionally, we provided a clear distinction between runtime and non-runtime Python typing concepts, and offered a clearer perspective on this using \emph{subclass-of}, \emph{object-instance-of}, and \emph{type-instance-of} relationships. Finally, we established a formal typing foundation using abstract data types and existential types, with which we demonstrated that a cohesive and elegant description of Python's type system is possible.

The formal foundation established in this paper is merely the first step, albeit a crucial one. Our future work will proceed along two main paths:
\begin{itemize}[leftmargin=*]
    \item \emph{Development of a type inference framework}: the primary motivation for this research is to build a sound static type inference tool. Our next step is to leverage the ADT-based formalisms to design and implement a type inference engine that is able to compute specifications of classes in isolation. We will use our previous work~\cite{spytype}\cite{nacujlamp} as the foundation for the practical implementation.

    \item \emph{Extension and refinement of the formalism}: one of the main directions for the extension of the formalism is related to finding a definition for \emph{ADT Subtyping}. We plan on working on the foundation provided by the covariant subtyping rule, also mentioned in Cardelli and Wegner's work~\cite{cardelli}. Also, we pay attention that Python has many complex and dynamic concepts to be addressed. For example, decorator patterns that alter function or class signatures and the ever-growing number of accepted PEP proposals.
\end{itemize}
However, there are multiple other avenues that warrant exploration. For example, the programatic extraction of formal ADT expressions from stub files, or a formal proof of soundness for a well-defined subset of the Python language to provide mathematical guarantess of type safety.

\bibliographystyle{eptcs}
\bibliography{from2025}

\begin{thebibliography}{10}
\providecommand{\bibitemdeclare}[2]{}
\providecommand{\surnamestart}{}
\providecommand{\surnameend}{}
\providecommand{\urlprefix}{Available at }
\providecommand{\url}[1]{\texttt{#1}}
\providecommand{\href}[2]{\texttt{#2}}
\providecommand{\urlalt}[2]{\href{#1}{#2}}
\providecommand{\doi}[1]{doi:\urlalt{https://doi.org/#1}{#1}}
\providecommand{\eprint}[1]{arXiv:\urlalt{https://arxiv.org/abs/#1}{#1}}
\providecommand{\bibinfo}[2]{#2}

\bibitemdeclare{article}{c3paper}
\bibitem{c3paper}
\bibinfo{author}{Kim \surnamestart Barrett\surnameend}, \bibinfo{author}{Bob
  \surnamestart Cassels\surnameend}, \bibinfo{author}{Paul \surnamestart
  Haahr\surnameend}, \bibinfo{author}{David~A. \surnamestart Moon\surnameend},
  \bibinfo{author}{Keith \surnamestart Playford\surnameend} \&
  \bibinfo{author}{P.~Tucker \surnamestart Withington\surnameend}
  (\bibinfo{year}{1996}): \emph{\bibinfo{title}{A monotonic superclass
  linearization for Dylan}}.
\newblock {\slshape \bibinfo{journal}{SIGPLAN Not.}}
  \bibinfo{volume}{31}(\bibinfo{number}{10}), p. \bibinfo{pages}{69–82},
  \doi{10.1145/236338.236343}.

\bibitemdeclare{inproceedings}{cardellisemantics}
\bibitem{cardellisemantics}
\bibinfo{author}{Luca \surnamestart Cardelli\surnameend}
  (\bibinfo{year}{1984}): \emph{\bibinfo{title}{A semantics of multiple
  inheritance}}.
\newblock In \bibinfo{editor}{Gilles \surnamestart Kahn\surnameend},
  \bibinfo{editor}{David~B. \surnamestart MacQueen\surnameend} \&
  \bibinfo{editor}{Gordon \surnamestart Plotkin\surnameend}, editors: {\slshape
  \bibinfo{booktitle}{Semantics of Data Types}}, \bibinfo{publisher}{Springer
  Berlin Heidelberg}, \bibinfo{address}{Berlin, Heidelberg}, pp.
  \bibinfo{pages}{51--67}, \doi{10.1007/3-540-13346-1_2}.

\bibitemdeclare{article}{cardelli}
\bibitem{cardelli}
\bibinfo{author}{Luca \surnamestart Cardelli\surnameend} \&
  \bibinfo{author}{Peter \surnamestart Wegner\surnameend}
  (\bibinfo{year}{1985}): \emph{\bibinfo{title}{On understanding types, data
  abstraction, and polymorphism}}.
\newblock {\slshape \bibinfo{journal}{ACM Comput. Surv.}}
  \bibinfo{volume}{17}(\bibinfo{number}{4}), p. \bibinfo{pages}{471–523},
  \doi{10.1145/6041.6042}.

\bibitemdeclare{misc}{cpythongit}
\bibitem{cpythongit}
 (\bibinfo{year}{2025}): \emph{\bibinfo{title}{python/cpython: The Python
  programming language}}.
\newblock \bibinfo{howpublished}{\url{https://github.com/python/cpython}}.
\newblock \bibinfo{note}{Accessed: 2025-06-21}.

\bibitemdeclare{misc}{duckwiki}
\bibitem{duckwiki}
 (\bibinfo{year}{2013}): \emph{\bibinfo{title}{Duck typing - Wikipedia}}.
\newblock
  \bibinfo{howpublished}{\url{https://en.wikipedia.org/wiki/Duck_typing}}.
\newblock \bibinfo{note}{Accessed: 2025-06-14}.

\bibitemdeclare{article}{liskov1994}
\bibitem{liskov1994}
\bibinfo{author}{Barbara~H. \surnamestart Liskov\surnameend} \&
  \bibinfo{author}{Jeannette~M. \surnamestart Wing\surnameend}
  (\bibinfo{year}{1994}): \emph{\bibinfo{title}{A behavioral notion of
  subtyping}}.
\newblock {\slshape \bibinfo{journal}{ACM Trans. Program. Lang. Syst.}}
  \bibinfo{volume}{16}(\bibinfo{number}{6}), p. \bibinfo{pages}{1811–1841},
  \doi{10.1145/197320.197383}.

\bibitemdeclare{inproceedings}{mitchellexistential}
\bibitem{mitchellexistential}
\bibinfo{author}{John~C. \surnamestart Mitchell\surnameend} \&
  \bibinfo{author}{Gordon~D. \surnamestart Plotkin\surnameend}
  (\bibinfo{year}{1985}): \emph{\bibinfo{title}{Abstract Types Have Existential
  Type}}.
\newblock In: {\slshape \bibinfo{booktitle}{Proceedings of the 12th ACM
  SIGACT-SIGPLAN Symposium on Principles of Programming Languages}},
  \bibinfo{series}{POPL '85}, \bibinfo{publisher}{Association for Computing
  Machinery}, \bibinfo{address}{New York, NY, USA}, p.
  \bibinfo{pages}{37–51}, \doi{10.1145/318593.318606}.

\bibitemdeclare{inproceedings}{monatecoop}
\bibitem{monatecoop}
\bibinfo{author}{Rapha\"{e}l \surnamestart Monat\surnameend},
  \bibinfo{author}{Abdelraouf \surnamestart Ouadjaout\surnameend} \&
  \bibinfo{author}{Antoine \surnamestart Min\'{e}\surnameend}
  (\bibinfo{year}{2020}): \emph{\bibinfo{title}{{Static Type Analysis by
  Abstract Interpretation of Python Programs}}}.
\newblock In \bibinfo{editor}{Robert \surnamestart Hirschfeld\surnameend} \&
  \bibinfo{editor}{Tobias \surnamestart Pape\surnameend}, editors: {\slshape
  \bibinfo{booktitle}{34th European Conference on Object-Oriented Programming
  (ECOOP 2020)}}, {\slshape \bibinfo{series}{Leibniz International Proceedings
  in Informatics (LIPIcs)}} \bibinfo{volume}{166}, \bibinfo{publisher}{Schloss
  Dagstuhl -- Leibniz-Zentrum f{\"u}r Informatik}, \bibinfo{address}{Dagstuhl,
  Germany}, pp. \bibinfo{pages}{17:1--17:29},
  \doi{10.4230/LIPIcs.ECOOP.2020.17}.

\bibitemdeclare{misc}{mopsa}
\bibitem{mopsa}
 (\bibinfo{year}{2025}): \emph{\bibinfo{title}{MOPSA Project}}.
\newblock \bibinfo{howpublished}{\url{https://mopsa.lip6.fr/}}.
\newblock \bibinfo{note}{Accessed: 2025-06-21}.

\bibitemdeclare{misc}{mypy}
\bibitem{mypy}
 (\bibinfo{year}{2025}): \emph{\bibinfo{title}{mypy - Optional Static Typing
  for Python}}.
\newblock \bibinfo{howpublished}{\url{https://mypy-lang.org/}}.
\newblock \bibinfo{note}{Accessed: 2025-06-18}.

\bibitemdeclare{article}{nacujlamp}
\bibitem{nacujlamp}
\bibinfo{author}{Andrei \surnamestart Nacu\surnameend} (\bibinfo{year}{2025}):
  \emph{\bibinfo{title}{Towards a type-based abstract semantics for Python}}.
\newblock {\slshape \bibinfo{journal}{Journal of Logical and Algebraic Methods
  in Programming}} \bibinfo{volume}{143}, p. \bibinfo{pages}{101032},
  \doi{10.1016/j.jlamp.2024.101032}.
\newblock
  \urlprefix\url{https://www.sciencedirect.com/science/article/pii/S2352220824000865}.

\bibitemdeclare{misc}{pep3119}
\bibitem{pep3119}
 (\bibinfo{year}{2007}): \emph{\bibinfo{title}{PEP 3119 – Introducing
  Abstract Base Classes}}.
\newblock \bibinfo{howpublished}{\url{https://peps.python.org/pep-3119/}}.
\newblock \bibinfo{note}{Accessed: 2025-06-14}.

\bibitemdeclare{misc}{pep484}
\bibitem{pep484}
 (\bibinfo{year}{2014}): \emph{\bibinfo{title}{PEP 484 – Type Hints}}.
\newblock \bibinfo{howpublished}{\url{https://peps.python.org/pep-0484/}}.
\newblock \bibinfo{note}{Accessed: 2025-06-14}.

\bibitemdeclare{misc}{pep544}
\bibitem{pep544}
 (\bibinfo{year}{2017}): \emph{\bibinfo{title}{PEP 544 – Protocols:
  Structural subtyping (static duck typing)}}.
\newblock \bibinfo{howpublished}{\url{https://peps.python.org/pep-0544/}}.
\newblock \bibinfo{note}{Accessed: 2025-06-14}.

\bibitemdeclare{misc}{pep604}
\bibitem{pep604}
 (\bibinfo{year}{2019}): \emph{\bibinfo{title}{PEP 604 – Allow writing union
  types as X | Y | peps.python.org}}.
\newblock \bibinfo{howpublished}{\url{https://peps.python.org/pep-0604/}}.
\newblock \bibinfo{note}{Accessed: 2025-06-21}.

\bibitemdeclare{book}{pierce-tpl}
\bibitem{pierce-tpl}
\bibinfo{author}{Benjamin~C. \surnamestart Pierce\surnameend}
  (\bibinfo{year}{2002}): \emph{\bibinfo{title}{Types and programming
  languages}}.
\newblock \bibinfo{publisher}{{MIT} Press}.
\newblock
  \urlprefix\url{https://mitpress.mit.edu/9780262162098/types-and-programming-languages/}.

\bibitemdeclare{misc}{pyclasses}
\bibitem{pyclasses}
 (\bibinfo{year}{2025}): \emph{\bibinfo{title}{9. Classes — Python 3.13.7
  documentation}}.
\newblock
  \bibinfo{howpublished}{\url{https://docs.python.org/3/tutorial/classes.html}}.
\newblock \bibinfo{note}{Accessed: 2025-08-26}.

\bibitemdeclare{misc}{pydatamodel}
\bibitem{pydatamodel}
 (\bibinfo{year}{2025}): \emph{\bibinfo{title}{3. Data model — Python 3.13.5
  documentation}}.
\newblock
  \bibinfo{howpublished}{\url{https://docs.python.org/3/reference/datamodel.html}}.
\newblock \bibinfo{note}{Accessed: 2025-06-21}.

\bibitemdeclare{misc}{pylance}
\bibitem{pylance}
 (\bibinfo{year}{2025}): \emph{\bibinfo{title}{microsoft/pylance-release:
  Documentation and issues for Pylance}}.
\newblock
  \bibinfo{howpublished}{\url{https://github.com/microsoft/pylance-release}}.
\newblock \bibinfo{note}{Accessed: 2025-06-21}.

\bibitemdeclare{misc}{pyright}
\bibitem{pyright}
 (\bibinfo{year}{2025}): \emph{\bibinfo{title}{microsoft/pyright: Static Type
  Checker for Python}}.
\newblock \bibinfo{howpublished}{\url{https://github.com/microsoft/pyright}}.
\newblock \bibinfo{note}{Accessed: 2025-06-21}.

\bibitemdeclare{misc}{pystdunions}
\bibitem{pystdunions}
 (\bibinfo{year}{2019}): \emph{\bibinfo{title}{Built-in Types — Python 3.13.2
  documentation}}.
\newblock
  \bibinfo{howpublished}{\url{https://docs.python.org/3/library/stdtypes.html\#types-union}}.
\newblock \bibinfo{note}{Accessed: 2025-06-21}.

\bibitemdeclare{misc}{python23c3}
\bibitem{python23c3}
 (\bibinfo{year}{2003}): \emph{\bibinfo{title}{The Python 2.3 Method Resolution
  Order}}.
\newblock
  \bibinfo{howpublished}{\url{https://www.python.org/download/releases/2.3/mro/}}.
\newblock \bibinfo{note}{Accessed: 2025-06-14}.

\bibitemdeclare{misc}{pyunif}
\bibitem{pyunif}
 (\bibinfo{year}{2003}): \emph{\bibinfo{title}{Unifying types and classes in
  Python 2.2}}.
\newblock
  \bibinfo{howpublished}{\url{https://www.python.org/download/releases/2.2.3/descrintro/}}.
\newblock \bibinfo{note}{Accessed: 2025-06-14}.

\bibitemdeclare{book}{ramalho2015fluent}
\bibitem{ramalho2015fluent}
\bibinfo{author}{Luciano \surnamestart Ramalho\surnameend}
  (\bibinfo{year}{2015}): \emph{\bibinfo{title}{Fluent Python: Clear, Concise,
  and Effective Programming}}.
\newblock \bibinfo{publisher}{O'Reilly Media}.
\newblock
  \urlprefix\url{https://www.oreilly.com/library/view/fluent-python/9781491946237/}.

\bibitemdeclare{misc}{spytype}
\bibitem{spytype}
 (\bibinfo{year}{2024}): \emph{\bibinfo{title}{andreinaku/SpyType}}.
\newblock \bibinfo{howpublished}{\url{https://github.com/andreinaku/SpyType}}.
\newblock \bibinfo{note}{Accessed: 2025-06-21}.

\bibitemdeclare{misc}{typingconcepts}
\bibitem{typingconcepts}
 (\bibinfo{year}{2025}): \emph{\bibinfo{title}{Type system concepts — typing
  documentation}}.
\newblock
  \bibinfo{howpublished}{\url{https://typing.python.org/en/latest/spec/concepts.html}}.
\newblock \bibinfo{note}{Accessed: 2025-08-21}.

\bibitemdeclare{misc}{virtualcpp}
\bibitem{virtualcpp}
 (\bibinfo{year}{2025}): \emph{\bibinfo{title}{Derived classes -
  cppreference.com}}.
\newblock
  \bibinfo{howpublished}{\url{https://en.cppreference.com/w/cpp/language/derived_class.html}}.
\newblock \bibinfo{note}{Accessed: 2025-06-16}.

\end{thebibliography}
\end{document}